\documentclass[11pt]{article}
\usepackage{graphicx}
\usepackage{amsmath}

\begin{document}

\def\xslash#1{{\rlap{$#1$}/}}
\def \p {\partial}
\def \dd {\psi_{u\bar dg}}
\def \ddp {\psi_{u\bar dgg}}
\def \pq {\psi_{u\bar d\bar uu}}
\def \jpsi {J/\psi}
\def \psip {\psi^\prime}
\def \to {\rightarrow}
\def\bfsig{\mbox{\boldmath$\sigma$}}
\def\DT{\mbox{\boldmath$\Delta_T $}}
\def\xit{\mbox{\boldmath$\xi_\perp $}}
\def \jpsi {J/\psi}
\def\bfej{\mbox{\boldmath$\varepsilon$}}
\def \t {\tilde}
\def\epn {\varepsilon}
\def \up {\uparrow}
\def \dn {\downarrow}
\def \da {\dagger}
\def \pn3 {\phi_{u\bar d g}}

\def \p4n {\phi_{u\bar d gg}}

\def \bx {\bar x}
\def \by {\bar y}

\begin{center} 
%{\Large\bf   Revisiting the Axial Gauge Puzzle  }

\par
{\Large\bf   Parton Distributions of Gluons in Different Representations} 
\par\vskip20pt
 J.P. Ma$^{1,2,3}$ and G.P. Zhang$^{4}$    \\
{\small {\it
$^1$ CAS Key Laboratory of Theoretical Physics, Institute of Theoretical Physics, P.O. Box 2735, Chinese Academy of Sciences, Beijing 100190, China\\
$^2$ School of Physical Sciences, University of Chinese Academy of Sciences, Beijing 100049, China\\
$^3$ School of Physics and Center for High-Energy Physics, Peking University, Beijing 100871, China\\
$^4$ Department of Physics, Yunnan University, Kunming, Yunnan 650091, China}} \\
\end{center}

\vskip 1cm
\begin{abstract}
It has been questioned if gluon distributions defined in different representations of $SU(N)$ are the same. The question has arisen in the connection to the recently proposed resolution of the so-called axial gauge puzzle. We give a proof that gluon distributions defined in the fundamental- and adjoint representation, where gauge links 
consist of those which are all future-pointing or all past-pointing, are same. These gauge links are needed to make the distributions gauge invariant. When these links consist of not only future-pointing gauge links but also 
past-pointing ones, then the puzzle appears. We examine the puzzle for distributions  
at one-loop in the cases with gauge links along time-like-, space-like- and light-cone direction.

\end{abstract}      
\vskip 5mm

\vskip40pt

Correlation functions defined with two field operators implemented with gauge links play important role in high 
energy physics, e.g., collinear parton distributions\cite{DFPDF}, transverse-momentum-dependent parton distributions\cite{CSS,JMY} and generalized parton distributions\cite{DMGPD,DVCSJi}.  
These distributions are gauge invariant and contain informations about inner structure of hadrons. 
Knowing them,  predictions about high energy scattering can be made with QCD factorization theorems. 

\par 
In the aforementioned distributions the gauge links are along light-cone- or near light-cone directions. 
In describing  heavy quark or heavy quarkonium dynamics in the quark-gluon plasma,  heavy quark transport coefficient\cite{TCHQ} and heavy quarkonium transport coefficient\cite{TCHQQ} 
are introduced with gauge links along the time-direction.  The operator used to define heavy quark transport coefficient is: 
\begin{equation} 
{\mathcal O}_E^Q (t) = g_s^2 {\rm Tr} \biggr [ U^\dagger (t, -\infty) G^{0i} (t e_t) U^\dagger (\infty, t) U(\infty, 0) G^{0i} (0)  U(0,-\infty) \biggr ],
\end{equation}  
where $G^{\mu\nu}$ is gluon field strength tensor. The gauge links are defined along the time-direction $ e_t =(1,0,0,0)$:
\begin{equation} 
 U(t,-\infty) =P \exp \biggr \{- ig_s \int_{-\infty}^t d\lambda G^0 (\lambda e_t) \biggr \}, 
 \quad U(\infty, t ) =P \exp \biggr \{- ig_s \int_{t}^\infty d\lambda G^0 (\lambda e_t) \biggr \}.  
 \end{equation}  
In ${\mathcal O}_E^Q$ all operators are separated in time-direction and in the fundamental representation. 
The heavy quarkonium transport coefficient is defined with the operator:
\begin{equation} 
{\mathcal O}_E^{Q\bar Q}  (t) = \frac{1}{2} g_s^2 G^{a, 0i} (t e_t)  \biggr [ U^\dagger (\infty, t) U(\infty, 0) \biggr ]^{ab} G^{b, 0i} (0), 
\end{equation}  
where gauge links are defined in the adjoint representation. Here and in the following, matrices with indices give by Greek's letters are always in the adjoint representation. The two operators are defined in an arbitrary gauge
and are gauge invariant. If we take the axial gauge $e_t\cdot G=G^0=0$, the two operators are the same.  When one calculates the vacuum expectation values of the two operators in the axial gauge, 
they should be the same. However, they are not the same when they are calculated in Feynman gauge\cite{AGP0,AGP1,AGP2,AGP3}. This is the axial gauge puzzle.

 \par\vskip10pt
\begin{figure}[hbt]
\begin{center}
\includegraphics[width=14cm]{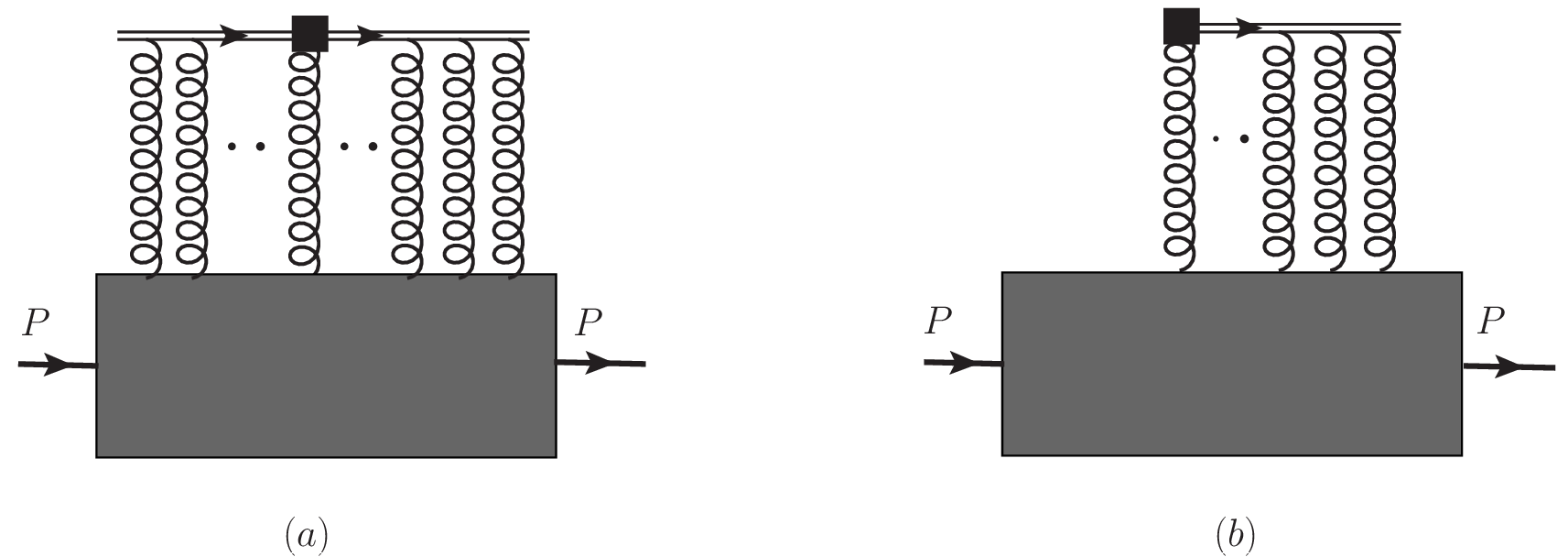}
\end{center}
\caption{ Two-gluon attachment to gauge links. The vertex marked as square is from the field 
strength tensor in the correlation functions. The vertices marked with circles are those from gauge links. 
 }
\label{Fig1p}
\end{figure}
\par 
An explanation or resolution for the puzzle is provided in \cite{BSXY}. In the quantization of gauge field theories by the path integral formalism a smooth gauge transformation from Feynman gauge to the axial gauge is essentially integral transformation.  It is shown in \cite{BSXY}  the Jacobian of the transformation for the zero mode of gauge fields $e_t\cdot G$ in Feynman gauge is zero. Therefore,  the transformation can not be performed properly.  A special gauge is introduced in \cite{BSXY},  which becomes under one limit Feynman gauge and under another limit the axial gauge. With this gauge the same difference in Feynman gauge is re-produced. 

After providing the resolution, the question is raised in \cite{BSXY}  if the identity  
\begin{equation} 
    U^\dagger (0,-\infty) G^{\mu\nu} (0)  U(0,-\infty) = T^a \biggr [ U^\dagger (0,-\infty) \biggr]^{ab} G^{b,\mu\nu} (0)
\label{IDT}    
\end{equation} 
holds at quantum level.  If it does not hold, a gluon distribution defined in the fundamental representation 
can be different than that defined in the adjoint representation. At classical level, it is relatively easy to prove the identity. At quantum level, it is difficult 
to prove because gauge fields as operators do not commute with each other.  It is noted that physical 
effects are not directly related to operators, but directly related to matrix elements of operators, or Green's functions.  In this letter, we show that the identity is true at quantum level in the sense that 
the operator in the left- and right-hand side of the identity appearing in Green's functions.  

The identity in the above is for gauge links along time-direction. In order to keep our proof more general, we introduce the gauge link along  an arbitrary direction given by $u$:
\begin{equation} 
 V_u (\infty, x) = P \exp\biggr \{ - ig_s \int_0^\infty d\lambda u\cdot G(\lambda u + x) \biggr \}. 
\end{equation} 
For our convenience we take the gauge link pointing to future. Correspondingly, we have the identity
\begin{equation} 
V_u (\infty, x)  G^{\rho \nu}  (x )  V_u^\dagger (\infty,x) = T^a  \left ( V_u (\infty,x) \right )_{ab}  G^{b, \rho \nu} (x ). 
\label{IDFA} 
\end{equation} 
To prove that the identity is true for Green's functions we introduce two correlation functions: 
\begin{eqnarray} 
 {\mathcal H}^{\nu} (x ) &=& \frac{ u_\rho} {u\cdot P} \int \frac{d\lambda}{2\pi} e^{-ix \lambda u\cdot P} 
 \langle P \vert  {\rm{Tr}} \biggr [ {\mathcal O}(\lambda u)  V_u (\infty,0)  G^{\rho \nu} (0 ) V_u^\dagger (\infty,0)  
\biggr ] \vert P \rangle, 
\nonumber\\
 {\mathcal G}^{\nu} (x ) &=& \frac{ u_\rho} {u\cdot P} \int \frac{d\lambda}{2\pi} e^{-ix \lambda u\cdot P} 
 \langle P \vert     {\rm Tr} \left [  {\mathcal O} (\lambda u) T^a \right ]   \left ( V_u (\infty,0) \right )_{ab}   G^{b, \rho \nu} (0 )  \vert P \rangle,   
\end{eqnarray}  
with ${\mathcal O} (x)$ as a gauge invariant operator. It is a matrix in the fundamental representation.
If the identity is true at quantum level in the sense discussed above, the two correlation functions are the same. In this letter, products of operators in a matrix element are always T-ordered products.   

\par
If we calculate the Green's function $ {\mathcal H}^{\nu}$ with perturbative theory, then contributions 
can be represented with Feynman diagrams.  Given a diagram contributing to $ {\mathcal H}^{\nu}$, we can always divide it into two parts. One part contains only the insertion of the operator $V_u (\infty,0)  G^{a,\rho \nu} (0 ) T^a V_u^\dagger (\infty,0)$ in the definition of $ {\mathcal H}^{\nu}$,  another part is the remainder. Two parts are connected with gluon lines, as shown in Fig.\ref{Fig1p}a. One gluon line is connected to the vertex represented by the small square, 
where the vertex stands for the insertion of $G^{a, \rho \nu} (0 )$. This gluon line carries the polarization index $\alpha$, the color index $a$, and the momentum $k$ flowing into the vertex. Other gluons lines are attached to the gauge links. For the diagram with the attachment of $n$-gluon lines to gauge links, the contribution 
can be written as:
\begin{eqnarray} 
 && \int d^4 k \left ( \prod_{i=1}^n d^4 k_i \right ) \biggr [  i  ( (k-\sum_{i=1}^n k_i)\cdot u  g^{\nu\alpha} - k^\nu u^\alpha ) \biggr ] u^{\mu_1}u^{\mu_2} \cdots u^{\mu_n} {\rm Tr}\biggr [  {\mathcal B}_n^{a a_1 a_2 \cdots a_n}  
\nonumber\\ 
 && \quad \quad  \Gamma^{aa_1a_2,\cdots a_n}_{\alpha\mu_1\mu_2\cdots \mu_n} (k, k_1,k_2,\cdots, k_n) \biggr ], 
 \label{HH}      
\end{eqnarray} 
$\Gamma$ stands for the grey or lower part in Fig.\ref{Fig1p}a, which emits $n+1$ gluons. Because of Bose-symmetry, 
$\Gamma$ is unchanged under an exchange between any two gluons. ${\mathcal B_n}$ combining 
the remaining factors represents the upper part with $n$-gluons attached to gauge links.  These $n$ gluons are labeled as $i$ with $i=1,2,\cdots, n$, the $i$-th gluon carries the polarization index $\mu_i$, the color index $a_i$, and the momentum $k_i$ flowing away from gauge links.

\begin{figure}[hbt]
\begin{center}
\includegraphics[width=9cm]{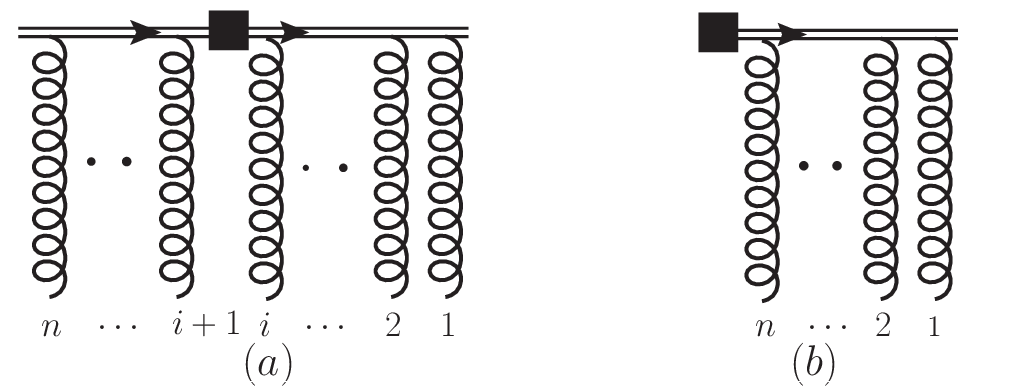}
\end{center}
\caption{ (a). Diagram for ${\mathcal A}_{i,n}$. There are $i$ gluons attached to the gauge link in the right side and $n-i$ gluons attached to that in the left side. The vertex marked as square is from the field 
strength tensor. (b). Diagram for $\tilde{\mathcal B}_n$ with the gauge link in the adjoint representation. 
 }
\label{Fig2p}
\end{figure}

Since there are two gauge links, among $n$-gluon there can be $i$ gluons attached to the gauge link in 
the right side of the square vertex, which represents $ V_u (\infty,0)$, and $n-i$ gluons are attached to the gauge link in the left side which is for $V_u^\dagger (0, \infty)$. Therefore ${\mathcal B}_n$ is the sum  
\begin{equation} 
    {\mathcal B_n} = \sum_{i=0}^n {\mathcal A}_{i.n}, 
\end{equation} 
where we have suppressed the indices $a, a_1,\cdots, a_n$.       
There are $n!$ possibilities or diagrams to attach $n$ gluons to gauge links. 
 Since the symmetry property of $\Gamma$,  each of these diagrams gives the same contribution after 
 the integration in Eq.(\ref{HH}).  Hence, we only need to consider any one of these diagrams.  
 We take the diagram in Fig.\ref{Fig2p}a for ${\mathcal A}_{i,n}$ which is given by: 
 \begin{eqnarray}
  {\mathcal A}_{i,n}  &=&  g_s^n (-1)^{n-i}  \frac{1}{u\cdot \bar k_1} \frac{1}{u\cdot \bar k_2} \cdots 
     \frac{1}{u\cdot \bar k_i } T^{a_1} T^{a_2} \cdots T^{a_i}  T^a T^{a_{i+1}} T^{a_{i+2}} \cdots T^{a_n}
\nonumber\\
  &&   \frac{1}{u\cdot (\bar k_n -\bar k_i )  } \frac{1}{u\cdot (\bar k_n-\bar k_{i+1}) }
  \cdots \frac{1}{u\cdot k_n} , 
  \end{eqnarray}
with the notation: 
\begin{equation} 
  \bar k_i =k_1+k_2+\cdots k_i, \quad  \bar k_n -\bar k_i = k_{i+1} + k_{i+2} + \cdots + k_n. 
   \end{equation}  
In the above, each eikonal propagator is associated with $+i\varepsilon$, e.g., $1/u\cdot \bar k_i$ is in fact
$1/(u\cdot \bar k_i + i\varepsilon)$. We have suppressed this factor. It is noted that the factor  $+i\varepsilon$ 
is from the fact that we have used gauge links pointing to the future. If gauge links point to the past, 
it should be $-i\varepsilon$.

\par 
Similarly, one has from Fig.\ref{Fig1p}b the contribution to ${\mathcal G}^\nu$ in the form 
\begin{eqnarray} 
 && \int d^4 k \left ( \prod_{i=1}^n d^4 k_i \right ) \biggr [  i  ( (k-\sum_{i=1}^n k_i)\cdot u  g^{\nu\alpha} - k^\nu u^\alpha ) \biggr ] u^{\mu_1}u^{\mu_2} \cdots u^n   \left (\tilde {\mathcal B}_n \right )_{ba}^{a_1a_2\cdots a_n}   
 \nonumber\\
   && \quad \quad {\rm Tr}\biggr [\Gamma^{aa_1a_2,\cdots a_n}_{\alpha\mu_1\mu_2\cdots \mu_n} (k, k_1,k_2,\cdots, k_n) T^b \biggr ], 
 \label{GG}      
\end{eqnarray} 
with $\tilde{\mathcal B}_n$ given by Fig.\ref{Fig2p}b as
 \begin{eqnarray}
   \left (\tilde {\mathcal B}_n \right )_{ba}  &=&  g_s^n  \frac{1}{u\cdot \bar k_1} \frac{1}{u\cdot \bar k_2} \cdots 
     \frac{1}{u\cdot \bar k_n } \biggr [ \hat T^{a_1} \hat T^{a_2} \cdots \hat T^{a_n} \biggr ]_{ba}.  
  \end{eqnarray}  
In the above $\hat T^a$ is the generator in the adjoint representation. It is given by:
\begin{equation} 
 \left ( \hat T^a\right )_{bc} = -i f^{abc} = -i f^{a}_{bc}, \quad \left [ T^a, T^b  \right ] =i f^{abc} T^c =  \left ( \hat T^a\right )_{cb} T^c
\end{equation} 
Using this relation, one has:
\begin{equation} 
\left [ T^{a_1}, \left [ T^{a_2}, \left [ \cdots, \left [ T^{a_n}, T^a\right ] \cdots \right ] \right ] \right ] 
  = \biggr [ \hat T^{a_1} \hat T^{a_2} \cdots \hat T^{a_n} \biggr ]_{ca} T^c. 
\end{equation} 

\par 
We will prove the identity in Eq.(\ref{IDFA}) by the induction in the below. We first consider the case $n=1$. With $n=1$ we have:
\begin{eqnarray} 
   {\mathcal B}_1^{a a_1}  = g_s \frac{1}{u\cdot k_1} T^{a_1} T^a - g_s \frac{1}{u\cdot k_1} T^{a} T^{a_1} =
                 g_s \frac{1}{u\cdot k_1}  T^c \left ( \hat T^c\right )_{a a_1}  = T^c \left (\tilde{\mathcal B}_1\right )^{a_1}_{ca} . 
\label{B1}                  
\end{eqnarray} 
This shows that in the case of $n=1$ the contribution from Fig.\ref{Fig1p}a and Fig.\ref{Fig1p}b are the same. 

\par 
In the case of $n=2$ we have: 
\begin{eqnarray} 
 {\mathcal B}_2^{a a_1 a_2} &=& g_s^2 \frac{1} {u\cdot k_1  }\frac{1}{u\cdot (k_1+  k_2) } T^{a_1} T^{a_2} T^{a}
   -g_s^2 \frac{1} {u\cdot k_1} T^{a_1} T^a T^{a_2} \frac{1}{u\cdot  k_2}
\nonumber\\
   && +g_s^2 T^{a} T^{a_1} T^{a_2 }\frac{1} {u\cdot (k_1+k_2) }\frac{1}{u\cdot  k_2}. 
 \end{eqnarray}
 We use the identity 
 \begin{equation} 
 \frac{1}{u\cdot k_1} \frac{1}{u\cdot k_2} =\frac{1}{u\cdot k_1} \frac{1}{u\cdot (k_1+k_2)} 
     +\frac{1}{u\cdot k_2} \frac{1}{u\cdot (k_1+k_2)}, 
 \end{equation} 
 to split the second term into two terms. If we make an exchange of $1\leftrightarrow 2$ for any term 
 in the above,  it will not affect the integral in Eq.(\ref{HH}). We can use this to  re-write $ {\mathcal B}_1^{a a_1 a_2}$ in the integration so that all terms have the same denominators. We have then:
\begin{eqnarray} 
 {\mathcal B}_2^{a a_1 a_2} &=& g_s^2 \frac{1} {u\cdot k_1  }\frac{1}{u\cdot (k_1+  k_2) }  \biggr [ 
 T^{a_1} T^{a_2} T^{a} - T^{a_1} T^a T^{a_2} - T^{a_1} T^a T^{a_2} +T^{a} T^{a_1} T^{a_2 } \biggr ]   
\nonumber\\
   &= & g_s^2  \frac{1} {u\cdot k_1  }\frac{1}{u\cdot (k_1+  k_2) } \left [ T^{a_1}, \left [ T^{a_2}, T^a\right ] \right ] = \left (\tilde{\mathcal B}_2\right )^{a_1a_2 }_{ca} T^c. 
 \end{eqnarray}  
 Hence, for $n=2$ the two contributions in Fig.\ref{Fig1p} are the same.  With little effort one can also shown
 that it is true for $n=3,4$. 
 
 Now we assume that with $n$-gluons the following equation is correct:
 \begin{equation} 
 {\mathcal B}_n^{a a_1a_2\cdots a_n}  =g_s^n  \frac{1}{u\cdot \bar k_1} \frac{1}{u\cdot \bar k_2} \cdots 
     \frac{1}{u\cdot \bar k_n } \left [ T^{a_1}, \left [ T^{a_2}, \left [ \cdots, \left [ T^{a_n}, T^a\right ] \cdots \right ] \right ] \right ] = T^c \left ( \tilde{\mathcal B}\right )^{a_1 a_2\cdots a_n}_{ca}, 
\label{NG}
\end{equation}       
and try to show that it also holds for $n+1$-gluons.  The contribution with $n+1$-gluons is
 \begin{equation} 
    {\mathcal B}_{n+1} = \sum_{i=0}^{n+1} {\mathcal A}_{i, n+1},
\end{equation}    
where ${\mathcal A}_{i, n+1}$ is given by Fig.\ref{Fig2p}a by attaching the additional  $n+1$-th gluon line to the two gauge links. For $i\neq n+1$  one can insert the $n+1$-th gluon in any place to attach the left gauge link. Different insertions give the same contribution because of the Bose symmetry of $\Gamma$.  
 Hence, we can write ${\mathcal A}_{i, n+1}$ with $i\neq n+1$ under the integrations as the contribution with 
 the extra gluon attached to the left gauge link nearest to the square vertex:   
\begin{eqnarray} 
  {\mathcal A}_{i, n+1}  &=&  g_s^{n+1} (-1)^{n+1-i}  \frac{1}{u\cdot \bar k_1} \frac{1}{u\cdot \bar k_2} \cdots 
     \frac{1}{u\cdot \bar k_{i-1}  } T^{a_1} T^{a_2} \cdots T^{a_{i-1}} \biggr [  \frac{1}{u\cdot \bar k_i }T^{a_i}  T^a 
     T^{a_{n+1}}  \frac{1}{u\cdot (\bar k_{n+1}  -\bar k_i )} \biggr ] 
 \nonumber\\
  &&   T^{a_{i+1}} T^{a_{i+2}} \cdots T^{a_n} \frac{1}{u\cdot (\bar k_n -\bar k_i )  } \frac{1}{u\cdot (\bar k_n-\bar k_{i+1}) }
  \cdots \frac{1}{u\cdot k_n}. 
  \end{eqnarray}   
We can write the term in $[\cdots ]$ in the first line, denoted as $B_2$ as:
\begin{eqnarray} 
 B_2 = T^{a_i} T^a T^{a_{n+1}} \frac{1}{u\cdot \bar k_i} \frac{1}{u\cdot ( \bar k_{n+1}  -\bar k_i ) } 
  =T^{a_i} T^a T^{a_{n+1}}\biggr [ \frac{1}{u\cdot \bar k_{n+1}} \frac{1}{u\cdot \bar k_i} 
 + \frac{1}{u\cdot \bar k_{n+1}} \frac{1}{u\cdot ( \bar k_{n+1}  -\bar k_i ) } \biggr ].  
\end{eqnarray}   
It is noted that the dependence on $k_i$ and $k_{n+1}$ in ${\mathcal A}_{i, n+1} $ is only in the above term. 
Again, because the Bose-symmetry we can make the exchange $i\leftrightarrow n+1$ in the second term in $B_2$, then we have: 
\begin{eqnarray} 
 B_2 = T^{a_i} T^a T^{a_{n+1}} \frac{1}{u\cdot \bar k_{n+1}} \frac{1}{u\cdot \bar k_i} 
 + T^{a_{n+1}} T^a T^{a_i }\frac{1}{u\cdot \bar k_{n+1}} \frac{1}{u\cdot ( \bar k_n -\bar k_{i-1} ) } .  
\end{eqnarray}      
With this decomposition ${\mathcal A}_{i, n+1}$ with $i\neq n+1$ is split into two terms: 
\begin{eqnarray} 
{\mathcal A}_{i, n+1} & =& {\mathcal C }_{i, n+1} +{\mathcal D}_{i, n+1},    
\nonumber\\
  {\mathcal C}_{i, n+1}  &=&  g_s^{n+1} (-1)^{n+1-i}  \frac{1}{u\cdot \bar k_1} \frac{1}{u\cdot \bar k_2} \cdots 
     \frac{1}{u\cdot \bar k_{i-1}  } T^{a_1} T^{a_2} \cdots T^{a_{i-1}} T^{a_i} T^a T^{a_{n+1}} \frac{1}{u\cdot \bar k_{n+1}} \frac{1}{u\cdot \bar k_i} 
 \nonumber\\
  &&   T^{a_{i+1}} T^{a_{i+2}} \cdots T^{a_n} \frac{1}{u\cdot (\bar k_n -\bar k_i )  } \frac{1}{u\cdot (\bar k_n-\bar k_{i+1}) }
  \cdots \frac{1}{u\cdot k_n} ,
\nonumber\\
  {\mathcal D}_{i, n+1}  &=&  g_s^{n+1} (-1)^{n+1-i}  \frac{1}{u\cdot \bar k_1} \frac{1}{u\cdot \bar k_2} \cdots 
     \frac{1}{u\cdot \bar k_{i-1}  } T^{a_1} T^{a_2} \cdots T^{a_{i-1}} T^{a_{n+1}} T^a T^{a_i }\frac{1}{u\cdot \bar k_{n+1}} \frac{1}{u\cdot ( \bar k_n -\bar k_{i-1} ) }
 \nonumber\\
  &&   T^{a_{i+1}} T^{a_{i+2}} \cdots T^{a_n} \frac{1}{u\cdot (\bar k_n -\bar k_i )  } \frac{1}{u\cdot (\bar k_n-\bar k_{i+1}) }
  \cdots \frac{1}{u\cdot k_n} ,
  \end{eqnarray} 
 and we have the following sum:
 \begin{eqnarray}
 \sum_{i=1}^{n-1} {\mathcal C}_{i, n+1} + \sum_{i=2}^n  {\mathcal D}_{i, n+1} &=& 
\frac{(-g_s)^{n+1} }{u\cdot \bar k_{n+1}} \sum_{i=1}^{n-1} (-1)^{-i}  \frac{1}{u\cdot \bar k_1} \frac{1}{u\cdot \bar k_2} \cdots 
     \frac{1}{u\cdot \bar k_{i-1}  }  \frac{1}{u\cdot \bar k_i} T^{a_1} T^{a_2} \cdots T^{a_{i-1}} T^{a_i}       
 \nonumber\\
  &&\biggr [ T^a , T^{a_{n+1}}  \biggr ]   T^{a_{i+1}} T^{a_{i+2}} \cdots T^{a_n} \frac{1}{u\cdot (\bar k_n -\bar k_i )  } \frac{1}{u\cdot (\bar k_n-\bar k_{i+1}) }
  \cdots \frac{1}{u\cdot k_n} ,
\label{PSUM}   
\end{eqnarray} 
The total contribution is
\begin{eqnarray}
 {\mathcal B}_{n+1} = \sum_{i=0}^{n+1} {\mathcal A}_{i, n+1}
= \sum_{i=1}^{n-1} {\mathcal C}_{i, n+1} + \sum_{i=2}^n  {\mathcal D}_{i, n+1} +\left ({\mathcal C}_{n, n+1} 
        +{\mathcal A}_{n+1, n+1}\right )   + \left ( {\mathcal D}_{1, n+1}+{\mathcal A}_{0, n+1} \right ) , 
\end{eqnarray}         
Using the above result it is easy to find that the last two terms in the second $(\cdots)$   give the $i=0$ term in the sum of the right side in Eq.(\ref{PSUM}), while the two terms in the first  $(\cdots)$ give the 
$i=n$ term. Hence we have: 
\begin{eqnarray} 
 {\mathcal B}_{n+1} &=& \sum_{i=0}^{n} \frac{(-g_s)^{n+1} }{u\cdot \bar k_{n+1}} (-1)^{-i}  \frac{1}{u\cdot \bar k_1} \frac{1}{u\cdot \bar k_2} \cdots 
     \frac{1}{u\cdot \bar k_{i-1}  }  \frac{1}{u\cdot \bar k_i} T^{a_1} T^{a_2} \cdots T^{a_{i-1}} T^{a_i}       
 \nonumber\\
  &&\biggr [ T^a , T^{a_{n+1}}  \biggr ]    T^{a_{i+1}} T^{a_{i+2}} \cdots T^{a_n} \frac{1}{u\cdot (\bar k_n -\bar k_i )  } \frac{1}{u\cdot (\bar k_n-\bar k_{i+1}) }
  \cdots \frac{1}{u\cdot k_n} ,
 \end{eqnarray}  
The sum of $n+1$ gluon attachment, i.e., ${\mathcal B}_{n+1}$ is obtained from ${\mathcal B}_n$ simply by the replacement 
in ${\mathcal B}_n$:
\begin{equation} 
  T^a \to \frac{   g_s}{u\cdot \bar k_{n+1} } [T^{a_{n+1}},T^a ] . 
\end{equation} 
Therefore, if Eq.(\ref{NG}) holds for the case with a given $n$, then it holds also for the case with $n+1$. By the meaning of the induction, we have proven that the two contributions 
in Fig.\ref{Fig1p} are the same for an arbitrary $n$.  This leads to the conclusion by summing over $n$ 
that  Eq,(\ref{IDT},\ref{IDFA}) are correct for Green's functions at quantum level.

With gauge links one can define two gluon distributions for a hadron state with the momentum $P^\mu =(P^0,0,0, P^3)$:
\begin{eqnarray}
 {\mathcal H}^{\mu\nu} (x ) &=& \frac{u_\sigma u_\rho} {u\cdot P} \int \frac{d\lambda}{2\pi} e^{-ix \lambda u\cdot P} 
 \langle P \vert  {\rm{Tr}} \biggr [ V_u (\infty, \lambda u )   G^{\sigma \mu} (\lambda u)  
V_u^\dagger (\infty,\lambda u) V_u (\infty,0)  G^{\rho \nu} (0 ) V_u^\dagger (\infty,0)  
\biggr ] \vert P \rangle,  
\nonumber\\
  {\mathcal G}^{\mu\nu} (x ) &=& \frac{u_\sigma u_\rho} {2 u\cdot P} \int \frac{d\lambda}{2\pi} e^{-ix \lambda u\cdot P} 
 \langle P \vert  G^{a, \sigma \mu} (\lambda u)  
\left [ V_u^\dagger (\infty,\lambda u) V_u (\infty,0) \right ] _{ab}   G^{b, \rho \nu} (0 )  \vert P \rangle.   
\end{eqnarray}      
If one takes $u$ along a light-cone direction with $u^2=0$ so that one has $u\cdot P\to ( P^0+P^3)/\sqrt{2}$,  $ {\mathcal G}^{\mu\nu} (x )$ is the standard gluon distribution function\cite{DFPDF}, where $x$ is the momentum fraction.  With the proved identity in Eq.(\ref{IDFA}) for Green's functions, it is straightforward to find that the two gluon distributions defined in Eq.(30) are the same. Therefore, two distributions can equivalently be used 
for factorization.  This is also true for transverse-momentum-dependent gluon distributions. 

\par 
Now we consider the case where we replace the gauge link $V^\dagger (\infty, 0)$ in ${\mathcal H}^\nu$ 
with $V(0,-\infty)$ to obtain the correlation function: 
\begin{eqnarray}
  {\mathcal F}^{\nu} (x ) &=& \frac{ u_\rho} {u\cdot P} \int \frac{d\lambda}{2\pi} e^{-ix \lambda u\cdot P} 
 \langle P \vert  {\rm{Tr}} \biggr [ {\mathcal O}(\lambda u)  
 V_u (\infty, 0)  G^{\rho \nu} (0 ) V_u (0, -\infty)  
\biggr ] \vert P \rangle. 
\end{eqnarray} 
This correlation function also receives a contribution from Fig.\ref{Fig1p}a.  For $n$-gluon attachment to gauge links, the contribution takes the same form as that of  ${\mathcal H}^\nu$ in Eq.(\ref{HH}) but with 
different ${\mathcal B}_n^{a a_1 a_2 \cdots a_n}$, because a gauge link is pointing to the past.  
It is easy to find that $ {\mathcal F}^{\nu} (x )$ can not be the same as $ {\mathcal G}^{\nu} (x )$ and the zero mode of $u\cdot G$ gluon field is the reason for the difference. To show this we consider the case $n=1$.  
 For $n=1$ we have 
\begin{equation} 
   {\mathcal B}_1^{a a_1}  = g_s \frac{1}{u\cdot k_1+i\varepsilon } T^{a_1} T^a - g_s \frac{1}{u\cdot k_1 - i\varepsilon} T^{a_1} T^a. 
\end{equation}
Here, we have given the factor $i\varepsilon$ explicitly in eikonal propagators because  the factor has different signs in different places. 
The contribution can be re-written as:
\begin{equation}
  {\mathcal B}_1^{a a_1} =   g_s \frac{1}{u\cdot k_1+i\varepsilon}  T^c \left ( \hat T^c\right )_{a a_1} - 2\pi i g_s \delta (u\cdot k_1) T^{a_1} T^a  = T^c \left (\tilde{\mathcal B}_1\right )^{a_1}_{ca} -  2\pi i g_s \delta (u\cdot k_1) T^{a_1} T^a .  
\end{equation} 
Unlike the case $n=1$ for ${\mathcal H}^\nu$ in Eq.(\ref{B1}),  an additional term proportional to $\delta (u\cdot k_1)$ appears here. Therefore, there is a difference between ${\mathcal F}^{\nu} (x )$ and $ {\mathcal G}^{\nu} (x )$, the difference only receives contributions from the zero mode of the gauge field $u\cdot G$.

If we take the axial gauge $u\cdot G=0$,  one finds from the definitions of ${\mathcal F}^{\nu} (x )$ and ${\mathcal G}^{\nu} (x )$  that the two distributions are the same because the gauge links in the gauge are unit
matrices. However, if we make a gauge transformation  from a gauge like Feynman gauge to the axial gauge, a difference between the two distributions appears. The axial gauge can be achieved by the transformation  with $V_u(\infty, x)$.  Under the transformation  $V_u (\infty,0)  G^{\rho \nu} (0 ) V_u^\dagger (\infty,0)$ in ${\mathcal H}^{\nu} (x )$ becomes the field strength tensor $G_A^{\mu\nu}(0)$ in the axial gauge, but $V_u (\infty,0)  G^{\rho \nu} (0 ) V_u (0,-\infty)$ in ${\mathcal F}^{\nu} (x )$  
becomes $G_A^{\rho \nu} (0 ) V_u (\infty,-\infty)$, i.e., it is not only the field strength tensor of the axial gauge, a gauge link is involved.  Therefore, the obtained ${\mathcal F}^{\nu} (x )$ in the axial gauge 
is not that obtained simply by taking the gauge directly.  This implies that the two Green's functions are not  the same in general.   
Since the path of the gauge link is from $-\infty$ to $\infty$, zero modes are involved.

\begin{figure}[hbt]
\begin{center}
\includegraphics[width=12cm]{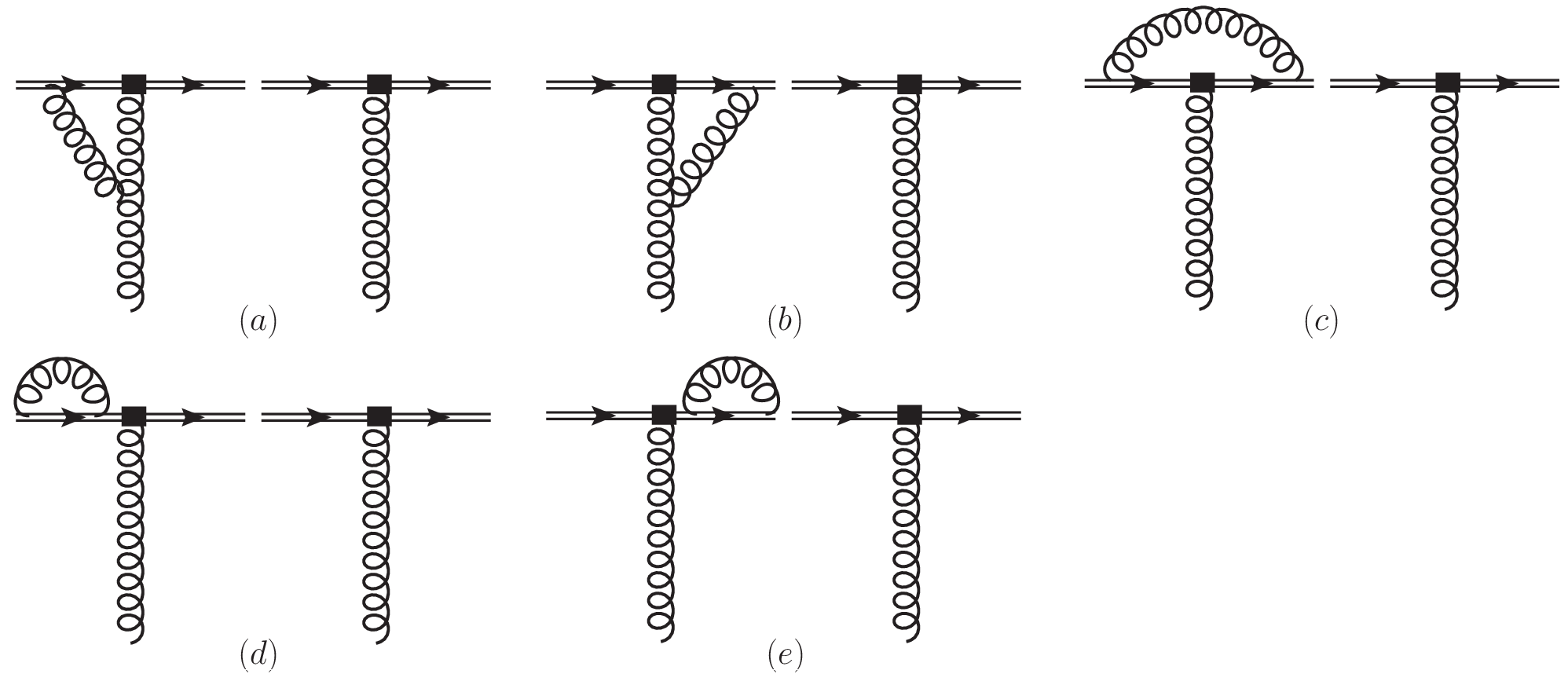}
\end{center}
\caption{  One-loop virtual correction to ${\mathcal H}^{\mu\nu}$ and ${\mathcal F}^{\mu\nu}$. }
\label{Fig3p}
\end{figure}

It is interesting to check the difference between $ {\mathcal H}^{\mu\nu} (x )$ and the following:
\begin{eqnarray}
  {\mathcal F}^{\mu\nu} (x ) = \frac{u_\sigma u_\rho} {u\cdot P} \int \frac{d\lambda}{2\pi} e^{-ix \lambda u\cdot P} 
 \langle P \vert  {\rm{Tr}} \biggr [ V_u^\dagger (\lambda u, -\infty )   G^{\sigma \mu} (\lambda u)  
V_u^\dagger (\infty,\lambda u ) V_u (\infty, 0)  G^{\rho \nu} (0 ) V_u (0, -\infty)  
\biggr ] \vert P \rangle
\end{eqnarray}      
by an explicit calculation. We will examine this with the state $\vert P\rangle$ as a gluon state with the momentum $k^\mu=(k^0, 0,0, k^3)$. Its color and spin are averaged. We take $u^\mu$ as $(u^0, 0,0,u^3)$. 
At tree-level, we have:
\begin{equation} 
{\mathcal H}^{\mu\nu} (x ) = {\mathcal F}^{\mu\nu} (x ) = -\frac{1}{4} \delta (1-x) g_\perp^{\mu\nu}, \quad 
g_\perp^{\mu\nu} = g^{\mu\nu} - \frac{u^\mu k^\nu + k^\mu u^\nu}{u\cdot k} 
   + u^2 \frac{k^\mu k^\nu}{(u\cdot k)^2}. 
\end{equation} 
There is no difference between two correlation functions at tree-level. The difference starts to be nonzero at one-loop. 
The difference is due to the different sign of $i\varepsilon$ factors in eikonal propagators originated from gauge links. One-loop correction has a virtual- and real part.  In the real part, the $i\varepsilon$ factor is irrelevant. The difference is then only from the virtual part. The relevant contributions are given by diagrams 
in Fig.\ref{Fig3p}, and their conjugated diagrams. 
\par 
The contribution from Fig.\ref{Fig3p}d is the self-energy correction of gauge links. With little algebra we find:
\begin{eqnarray} 
{\mathcal H}^{\mu\nu} (x )\biggr\vert_{3d} &=&\frac{1}{2} g_s^2 C_F \left ( -\frac{1}{4} \delta (1-x) g_\perp^{\mu\nu}\right ) 
\biggr [ - i u^2 \int \frac{d^4 k_1}{(2\pi)^4}  
     \frac{1}{- u\cdot k_1+i\varepsilon }  \frac{1}{u\cdot k_1+i\varepsilon}  
      \frac{1}{k_1^2+i\varepsilon}
       \biggr ],  
\nonumber\\
{\mathcal F}^{\mu\nu} (x ) \biggr\vert_{3d}&=&\frac{1}{2} g_s^2 C_F \left ( -\frac{1}{4} \delta (1-x) g_\perp^{\mu\nu}\right ) 
\biggr [ - i u^2 \int \frac{d^4 k_1}{(2\pi)^4}  
     \frac{1}{- u\cdot k_1- i\varepsilon }  \frac{1}{u\cdot k_1-i\varepsilon}  
      \frac{1}{k_1^2+i\varepsilon}
       \biggr ],  
\label{3D} 
\end{eqnarray}
where there is a pinch singularity in the loop integral which will be regularized later.  
The loop integral in the first line and the second line is the same. Therefore, the contribution from the self-energy to the difference is zero. The same applies for the contributions from Fig.\ref{Fig3p}e.

Before we turn to the contributions of the remaining diagrams, we briefly discuss how to regularize the loop 
integral with the pinch singularity in the above. It is noted that the $i\varepsilon$-factors in Eq.(\ref{3D}) have different origins. The factor 
in eikonal propagators comes from the path-ordering in gauge links, while that in the gluon propagator 
comes from the standard time-ordering of Green's functions. In principle the two factors can be different,  they are taken to be zero after integrations. Hence we can re-write the integral as:
\begin{equation} 
  I_p(\eta)  =\biggr [ -  \int \frac{d^d k_1}{(2\pi)^d}  
     \frac{1}{- u\cdot k_1+i\eta }  \frac{1}{u\cdot k_1+i\eta}  
      \frac{1}{k_1^2+i\varepsilon}
       \biggr ],       
\end{equation} 
where the pinch singularity is regularized with the finite and positive  $\eta$. Dimensional regularization with $d=4-\epsilon$ is used for the integration   After the integration one can take the limit $\eta\to 0^+$. 

With the above regularizations the contributions from Fig.\ref{Fig3p}c are:
\begin{eqnarray}
{\mathcal H}^{\mu\nu} (x )\biggr\vert_{3c} &=&-  \frac{i g_s^2 u^2}{4 N_c (d-2) } g_s^2 \delta (1-x) g_\perp^{\mu\nu}
   I_p(\eta) ,  
\nonumber\\
{\mathcal F}^{\mu\nu} (x )\biggr\vert_{3c} &=&-  \frac{i g_s^2 u^2}{4 N_c (d-2) } g_s^2 \delta (1-x) g_\perp^{\mu\nu} I_1 (\eta) ,  
\label{3C} 
\end{eqnarray}
with 
\begin{equation} 
I_1(\eta) = \int \frac{d^d k_1}{(2\pi)^d}  
     \frac{1}{(u\cdot k_1 -i\eta )^2 }  
      \frac{1}{k_1^2+i\varepsilon}. 
\end{equation}            
Performing the integration with the finite $\eta$ we find the following relation:
\begin{equation} 
    I_p (\eta ) =\frac{1}{d-3} I_1 (\eta)
\end{equation} 
for $u^2>0$ or $u^2<0$. In the limit $\eta\to 0^+$ we have:
\begin{equation}
   I_1 (0) = - \frac{i }{8\pi^2 u^2 } \biggr [ \frac{2}{\epsilon_{UV} }-\frac{2}{\epsilon_{IR} }\biggr ],  
\end{equation}        
where the first- and second pole in $d-4$ is for U.V.- and infrared divergence, respectively. 
The relation indicates that there is no difference between contributions from Fig.\ref{Fig3p}c to 
${\mathcal H}^{\mu\nu}$ and ${\mathcal F}^{\mu\nu}$ in the case with $u^2>0$ or $u^2<0$. 
For $u^2=0$ the contributions from Fig.\ref{Fig3p}c and the self-energy are zero. Therefore, the nonzero 
difference can only come from Fig.\ref{Fig3p}a and 3b and their complex conjugates.

The contribution to the difference from Fig.\ref{Fig3p}a after working out some trivial factors is:
\begin{eqnarray} 
  && {\mathcal F}^{\mu\nu} (x )\biggr\vert_{3a} - {\mathcal H}^{\mu\nu} (x )\biggr\vert_{3a}
 = -\frac{\pi}{2} g_s^2 N_c \frac{\delta (1-x)}{(u\cdot k)^2} 
 \int \frac{ d^d k_1}{(2\pi)^d} \frac{N^{\mu\nu} \delta (u\cdot k_1)}{(k_1^2 + i\varepsilon) ((k-k_1)^2 + i\varepsilon ) }, 
\nonumber\\
  && N^{\mu\nu}  =\frac{2}{(d-2) u\cdot k} \biggr [ \biggr ( g^{\mu\nu} - \frac{u^\mu u^\nu}{u^2} \biggr ) (u\cdot k)^4
      + u^2 \tilde k^\mu \tilde k^\nu \biggr ( (u\cdot k)^2 -u^2 k_1\cdot k \biggr ) 
        + \tilde k^\mu k_1^\nu (u^2)^2 k_1\cdot k 
\nonumber\\        
     && \quad \quad \quad   -(u\cdot k)^2 u^2 ( k_1^\mu \tilde k_1^\nu  -  k_1^\mu k_1^\nu )  \biggr ],  
     \quad \tilde k^\mu = k^\mu -u^\mu \frac{u\cdot k}{u^2}.  
\end{eqnarray} 
The contribution from Fig.\ref{Fig3p}b is the same as that from Fig.\ref{Fig3p}a. 
Inspecting the expression we need to calculate the three loop-integrals:
\begin{equation} 
 \left\{ I_0, I^{\mu}, I^{\mu\nu} \right \} =  \int \frac{ d^d k_1}{(2\pi)^d} \frac{\delta (u\cdot k_1)}{(k_1^2 + i\varepsilon) ((k-k_1)^2 + i\varepsilon ) }  \left\{ 1, k_1^{\mu},  k_1^\mu k_1^\nu  \right \}. 
\end{equation} 
To evaluate these integrals we distinguish the cases of $u^2>0$ and $u^2<0$. For simplicity we take in 
the time-like case  $u^\mu =(u^0,0,0,0)$, and in the space-like case we take $u^\mu =(0,0,0, u^3)$. In both cases we assume that $u\cdot k >0$. The time-like case corresponds to the case of correlation functions 
mentioned at beginning. The space-like case corresponds to the case of quasi gluon distributions\cite{Ji}. We obtain the result for $u^2>0$  
\begin{equation} 
 I_0 = 
    - \frac{(4\pi)^{\epsilon/2}}{16\pi^2 } (2 u\cdot k)^{-1-\epsilon}  \Gamma (1+\frac{\epsilon}{2})  \biggr [ i\frac{2}{\epsilon_{IR}}  - \pi +{\mathcal O}(\epsilon) \biggr ] , 
\end{equation} 
and for $u^2<0$:    
\begin{equation} 
 I_0=    - \frac{(4\pi)^{\epsilon/2}}{16\pi^2 } (2 u\cdot k)^{-1-\epsilon} \Gamma (1+\frac{\epsilon}{2} ) \biggr [ i\frac{2}{\epsilon_{IR}} +{\mathcal O}(\epsilon)  \biggr ] . 
\end{equation} 
The results for other two integrals with $u^2>0$ and $u^2<0$ are same. They are:
\begin{eqnarray} 
I^\mu = \frac{i}{16\pi^2 u\cdot k} \tilde k^\mu, 
\quad 
I^{\mu\nu} = -\frac{i}{32\pi^2} \frac{u\cdot k}{u^2} \biggr [ \biggr ( g^{\mu\nu} - \frac{u^\mu u^\nu}{u^2} \biggr ) 
     - \frac{u^2}{(u\cdot k)^2} \tilde k^\mu \tilde k^\nu \biggr ]. 
\end{eqnarray} 
From these results only $I_0$ with $u^2>0$ has a real part, while $I^0$ with $u^2<0$ and other two integrals 
have only  imaginary parts.  When adding the contribution from the conjugated diagram of Fig.\ref{Fig3p}a, 
only the contribution from the real part remains.

In the case of $u^2=0$, we take $u^\mu = (1,0,0,-1)/\sqrt{2}$.  Calculating the three loop integrals in this case,  we find 
\begin{equation} 
  I_0 = \frac{i}{32\pi^2 u\cdot k} \biggr [ \frac{2}{\epsilon_{UV}} -\frac{2}{\epsilon_{IR}}\biggr ] .
\end{equation}  
It is imaginary. Other two integrals are also imaginary. Unlike the case with $u^2>0$ or $u^2<0$, $I_0$ here is U.V. divergent. This indicates that one can not obtain correct results with $u^2=0$ from those with $u^2\neq 0$ by the limit $u^2\to 0$.

Based on the above results we conclude that the difference between the two correlation functions 
is only nonzero for $u^2>0$ at one-loop:
\begin{equation}
 {\mathcal F}^{\mu\nu} (x )- {\mathcal H}^{\mu\nu} (x ) = -\frac{g_s^2}{32} N_c \delta(1-x) \biggr [ 
    \biggr ( g^{\mu\nu}-\frac{u^\mu u^\nu}{u^2} \biggr ) + \frac{u^2}{(u\cdot k)^2} \tilde k^\mu \tilde k^\nu \biggr ]. 
\end{equation}     
Similarly, for transverse-momentum-dependent(TMD) gluon distributions, we obtain for $u^2>0$:
\begin{equation}
 {\mathcal F}^{\mu\nu} (x, p_\perp )- {\mathcal H}^{\mu\nu} (x,p_\perp ) = -\frac{g_s^2}{32} N_c \delta^2(p_\perp) \delta(1-x) \biggr [ 
    \biggr ( g^{\mu\nu}-\frac{u^\mu u^\nu}{u^2} \biggr ) + \frac{u^2}{(u\cdot k)^2} \tilde k^\mu \tilde k^\nu \biggr ], 
\end{equation}   
where TMD distributions are defined as
\begin{eqnarray}
  {\mathcal F}^{\mu\nu} (x, p_\perp ) &=& \frac{u_\sigma u_\rho} {u\cdot P} \int \frac{d\lambda}{2\pi} \frac{d^2 x_\perp}{(2\pi)^2} e^{-ix \lambda u\cdot P -i x_\perp\cdot p_\perp}  \langle P \vert  {\rm{Tr}} \biggr [ V_u^\dagger (\lambda u+x_\perp, -\infty )   G^{\sigma \mu} (\lambda u+x_\perp) 
\nonumber\\  
   && 
V_u^\dagger (\infty,\lambda u+x_\perp ) V_u (\infty, 0)  G^{\rho \nu} (0 ) V_u (0, -\infty)  
\biggr ] \vert P \rangle, 
\nonumber\\
 {\mathcal H}^{\mu\nu} (x, p_\perp ) &=& \frac{u_\sigma u_\rho} {u\cdot P} \int \frac{d\lambda}{2\pi} \frac{d^2 x_\perp }{(2\pi)^2 } e^{-ix \lambda u\cdot P -i x_\perp\cdot p_\perp}  \langle P \vert  {\rm{Tr}} \biggr [ V_u (\infty, \lambda u+x_\perp )   G^{\sigma \mu} (\lambda u+x_\perp) 
\nonumber\\  
   && 
V_u^\dagger (\infty,\lambda u+x_\perp ) V_u (\infty, 0)  G^{\rho \nu} (0 ) V_u^\dagger (\infty,0)  
\biggr ] \vert P \rangle,  
\end{eqnarray}    
with $x_\perp^\mu =(0,x_\perp^1, x_\perp^2, 0)$. The above definitions are given in a non-singular gauge 
like Feynman gauge, where the transverse gauge links at $\lambda u =\pm\infty$ can be omitted. 
For $u^2\leq 0$ the difference is zero at one loop.

It is interesting to note that the at one-loop two gluon distributions have a nonzero difference at one-loop only for $u^2>0$.  Although the 
gluon distributions with $u^2\leq 0$ have no difference at one-loop, but in general they are different. They will be different when the zero modes of the gauge field $u\cdot G$ contribute. 
 With small-$x$ approximation it is already shown that the two TMD distributions are different for $u^2=0$\cite{KKT,DXY,MZ}. 
But, after the integration over the transverse momentum, the difference disappears\cite{HTX}. 
% Summary 

In this letter,  we give a proof that gluon distributions defined in the fundamental- and adjoint representation with gauge links pointing to future are the same. This also applies for the case where all gauge links are pointing to past.  Gluon distributions defined only with future-pointing- or past-pointing gauge links are in general different than those defined with a combination of  future-pointing- and past-pointing gauge links, although they seem to be the same if one directly takes an axial gauge so that all gauge links become unit matrices.  However, if one makes a gauge transformation from a gauge into the axial gauge, one will see that there is a difference between two types of gluon distributions. We have taken a single-gluon state to calculate the difference at one loop. It is found that the difference for gauge links along time-like direction is not zero.  For space-like or light-cone directions the difference is zero.   

\par\vskip40pt
% The correct funding number:  12075299 (mian shang 2021-2024) 
% qunti 11821505, penghuanwu 11847612
% Adding lattice funding number 11935017(2020-2024)
% The number 11675241 is for 1017-2020. 

\noindent
{\bf Acknowledgments}
\par
J.P. Ma would like to thank Dr. J. Zhou for an interesting discussion. The work is supported by National Natural Science Foundation of P.R. China(No.12075299,11821505,11935017 and 12065024)  and by the Strategic Priority Research Program of Chinese Academy of Sciences, Grant No. XDB34000000.

\par

\end{document}